\documentclass[twocolumn,amsmath,amssymb]{revtex4}
\usepackage{epsfig}
\usepackage{color}
\newcommand{\highlight}{\color{black}}

\newcommand{\nnn}{\boldsymbol{n}}
\newcommand{\rrr}{\boldsymbol{r}}
\newcommand{\nablabf}{\boldsymbol{\nabla}}

\begin{document}

\title{\highlight Screening model for nanowire surface-charge sensors in liquid}

\author{Martin Hedeg\aa rd S\o rensen, Niels Asger Mortensen, and Mads Brandbyge}

\affiliation{MIC -- Department of Micro and Nanotechnology,
Nano$\bullet$DTU, Technical University of Denmark, DTU-building
345 east, DK-2800 Kongens Lyngby, Denmark}

\date{July 16, 2007}

\begin{abstract}
The conductance change of nanowire field-effect transistors is
considered a highly sensitive probe for surface charge. However,
Debye screening of relevant physiological liquid environments
challenge device performance due to competing screening from the
ionic liquid and nanowire charge carriers. We discuss this effect
within Thomas--Fermi and Debye--H{\"u}ckel theory and derive
analytical results for cylindrical wires which can be used to
estimate the sensitivity of nanowire surface-charge sensors. We
study the interplay between the nanowire radius, the Thomas--Fermi
and Debye screening lengths, and the length of the
functionalization molecules. The analytical results are compared
to finite-element calculations on a realistic geometry.
\end{abstract}

 \maketitle

Imagine a sensor so small and compact that it can fit almost
everywhere and detect all kinds of chemical substances in real
time. Such a sensor could in principle monitor and detect unwanted
bacteria and viruses instantaneously in eg. your blood or drinking
water. Sensors based on semiconductor nanowires have already been
fabricated~\cite{Cui:2001} and shown to work, for example as a pH
sensor, where the concentration of hydrogen ions H$^{+}$ in a
surrounding liquid is detected~\cite{Chen:2006}. Moreover,
applications in label-free detection and biological sensing
addressing e.g. DNA in low concentration is now being
explored~\cite{Zheng:2005b,Wang:2005,Stern:2007,Carlen:2007}. In
general, the conductance $G$ of a nanowire is in the literature
considered a promising candidate for a highly {\highlight
sensitive} probe of charged particles covering or situated near
its surface. Detection limits have been studied theoretically with
respect to binding-diffusion dynamics of the molecules at and near
the surface~\cite{Sheehan:2005,Nair:2006}. However, it is also
crucial to address the fundamental problem of effective screening
of these charges in physiologically relevant liquids which
typically yield screening of charges on the nanometer length
scale. Of course it is possible to decrease the salt concentration
in the analyte solution, but this may not only complicate the
process -- it may also change the biological functions.

\begin{figure}[b!]
\begin{center}
\epsfig{file=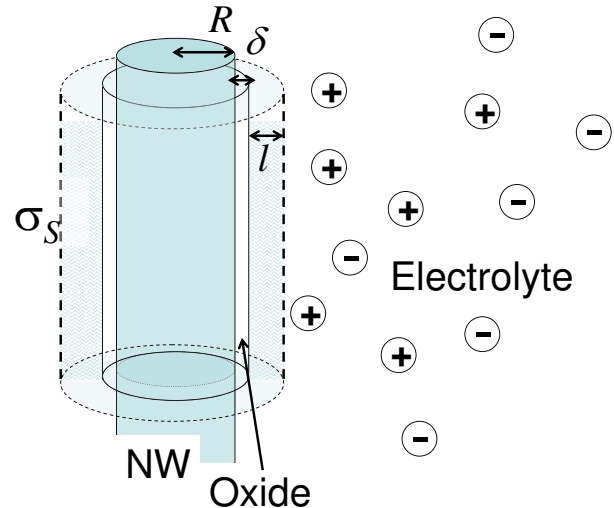, width=\columnwidth,clip}
\end{center}
\caption{Schematic cross section of a nanowire covered by an
ultra-thin oxide ($\delta$) immersed in an electrolyte. A positive
surface-charge density $\sigma_S$, supported by functionalization
molecules of length $\ell$, is screened by negative ions from the
electrolyte and electrons from the nanowire. For the opposite sign
of $\sigma_S$, the surface charge will be screened by attracting
positive ions from the electrolyte and by expelling the electrons
from the nearby-surface region of the nanowire.} \label{fig1}
\end{figure}

Effects of screening have been discussed qualitatively in the
literature (see e.g. Refs.~\onlinecite{Cheng:2006,Stern:2007})
touching on the influence from the competing screening mechanisms
of the electron gas in the nanowire and the dissolved ions in the
surrounding electrolyte. Very recently the effect has also been
discussed in a computational study of a Silicon nanowire (SiNW)
FET~\cite{Heitzinger:2007}. In this Letter we develop a simple
screening model which exemplifies how the desired screening of
surface charge by the nanowire charge carriers may be jeopardized
by the contending screening by the electrolyte. We derive
analytical results for cylindrical wires which can be used to
estimate the sensitivity of a nanowire surface-charge sensor
surrounded by electrolyte. The analytical results are compared to
finite-element calculations on a realistic geometry.

A surface charge density $\sigma_S$ will perturb the initial
charge-carrier density $n_0$ in the nanowire by $\Delta n$, thus
changing the initial conductance $G_0$ by $\Delta G$. As an
example, a positive surface charge will attract additional
electrons from the nanowire contacts (equilibrium reservoirs).
From simple conservation of charge we arrive at the following
expression for the sensitivity
{\highlight
\begin{equation}\label{eq:bound}
\frac{\Delta G}{G_0} = \frac{\sigma_S P}{ en_0 A }\times
\Gamma\propto n_0^{-2/3}
\end{equation} }
with $\Gamma$ being a dimensionless function, between zero and
unity, quantifying the actual sensitivity in the presence of Debye
screening in the electrolyte and a finite Thomas--Fermi screening
in the nanowire. {\highlight We show that $\Gamma$ scales with the
electron density $n_0$ as $n_0^{1/3}$ in the dilute limit, thus
leading to a prediction of a $n_0^{-2/3}$ dependence, contrasting
the intuitively expected $n_0^{-1}$ dependence. However, high
sensitivity is of course still associated with low densities.}

Above, $P$ is the perimeter of the of nanowire cross section (the
fraction supporting the surface charge), $A$ is the
cross-sectional area, and $e$ is the electron charge. For a
cylindrical wire of radius $R$ we have $P/A=2/R$, thus clearly
illustrating the benefit of scaling the wires to the nano regime.
{\highlight Obviously, screening in the liquid will suppress the
sensitivity below the bound given by Eq.~(\ref{eq:bound}) and a
$n_0^{-1}$ dependence (the limit $\Gamma=1$) can only be expected
when the Thomas--Fermi screening in the nanowire is much stronger
than the Debye screening in the electrolyte surrounding the
nanowire so that changes in the density of the electron gas fully
compensates the additional surface charge.} In our screening model
we consider the induced electrical potential due to a
surface-charge density $\sigma_S$ on the outside of an
oxide-covered nanowire, see Fig.~\ref{fig1}. For the nanowire we
employ the Thomas--Fermi model (see e.g.
Ref.~\onlinecite{Ashcroft:1976}) while for the dilute electrolyte
we consider a Debye--H\"uckel approximation (see e.g.
Ref.~\onlinecite{Feynman:1964b}). We these approximations we
arrive at the following linear differential equation for the
induced electrical potential $\phi$,
\begin{subequations}
\begin{equation}\label{eq:master}
\nabla^2\phi=\left\{ \begin{matrix} \lambda_{\rm
TF}^{-2}\:\phi&,&\rrr\in \Omega_{1},\\\\
0 &,& \rrr \in \Omega_{2},\\\\
\lambda_{D}^{-2}\:\phi&,&\rrr \in \Omega_{3}&\vee&
\Omega_{4},\end{matrix}\right.
\end{equation}
where $\lambda_{\rm TF}$ is the Thomas--Fermi screening length in
the nanowire domain $\Omega_{1}$ and $\lambda_{D}$ is the Debye
screening length in the electrolyte domains $\Omega_{3}$ and
$\Omega_{4}$. The electrically insulating oxide-layer domain
$\Omega_2$ is free from charges and the potential is simply a
solution to the Laplace equation. Obviously, the induced potential
should vanish at infinity and denoting the solution in domain
$\Omega_i$ by $\phi_i$ and the corresponding dielectric function
by $\epsilon_i$ we have the additional boundary conditions
\begin{equation}\label{eq:boundary}
\left.\begin{matrix}\phi_i-\phi_{i+1}=0\\\\
\nnn\cdot \nablabf \left[\epsilon_i\phi_i -
\epsilon_{i+1}\phi_{i+1}\right] =
\sigma_{i,i+1}\end{matrix}\right\},\quad
\rrr\in\partial\Omega_{i,i+1}
\end{equation}
\end{subequations}
where $\nnn$ is a normal vector to the surface
$\partial\Omega_{i,i+1}$ separating the neighboring domains
$\Omega_{i}$ and $\Omega_{i+1}$. Furthermore, $\sigma_{i,i+1}$ is
the corresponding surface-charge density. In the following we
consider the case of a surface-charge density $\sigma_S$
accumulated at the surface $\partial\Omega_{3,4}$, i.e.
$\sigma_{1,2}=\sigma_{2,3}=0$ and $\sigma_{3,4}=\sigma_S$. For the
dielectric function we have $\epsilon_3=\epsilon_4$. The thickness
$\ell$ of the domain $\Omega_3$ is physically interpreted as the
length of the functionalization molecules supporting the charge
which for $\ell=0$ resides directly on the outside of the
oxide-layer of the nanowire. In general the oxide layer will be
charged unless the analyte solution equals the isoelectric point
of the surface. Here we neglect this charge and assume that it may
just shift the value of the initial carrier concentration in the
wire.

In the following we consider the conductance $G=G_0+\Delta G$ of the nanowire of length
$L$ and in particular we focus on the conductance change $\Delta G$ due to a finite
surface-charge density $\sigma_S$. In terms of the electron mobility $\mu$ we have $G=L
^{-1}\int_{\Omega_1} d\rrr\, en(\rrr)\:\mu$ where $n=n_0+\Delta n$ is the electron density
and $e$ is the electron charge. We determine the induced charge-density, $n=n_0+\Delta n$,
from the Poisson equation,
\begin{equation}
-e\:\Delta n
(\rrr)=-\epsilon_1\nabla^2\phi_1(\rrr)=-\frac{\epsilon_1}{\lambda_{\rm
TF}^2}\phi_1(\rrr).
\end{equation}
Formally, the conductance change can now be expressed in terms of
the induced potential, i.e.
\begin{equation}\label{eq:Gintegral}
\frac{\Delta G}{G_0} =\frac{\epsilon_1}{\lambda_{\rm TF}^2e n_0
 } \frac{\int_{\Omega_1} d\rrr\, \phi(\rrr)}{\int_{\Omega_1} d\rrr},
\end{equation}
where $G_0=A en_0 \mu$ is the conductance in the absence of surface charge.

Equation~(\ref{eq:Gintegral}) is our general result for the
relative conductance change expressed in terms of an integral over
the surface-charge induced potential in the nanowire. Combined
with Eqs.~(\ref{eq:master}) and (\ref{eq:boundary}) it forms a
starting point for analytical solutions of simple geometries or
numerical solutions of more complicated geometries, e.g. by a
finite-element method.

Before turning to more complicated geometries we first consider
the case of a long cylindrical nanowire immersed in an infinite
volume of electrolyte (see Fig.~\ref{fig1}). The nanowire has a
radius $R$ with an oxide-layer of thickness $\delta$ and due to
the cylinder symmetry we may solve the problem analytically. For
simplicity we focus on the, for experiments, highly relevant limit
$\delta \ll R$ of an ultra-thin oxide layer. Furthermore, we
consider $\ell=0$ so that the charge resides on the outside of the
nanowire oxide layer. Solving the linear problem we then arrive at
\begin{equation}\label{eq:Gamma_circle}
\Gamma\simeq  \left[1+ \frac{\epsilon_3}{\epsilon_1}
\frac{\lambda_{\rm TF}}{\lambda_D}\frac{
   I_0\left(\frac{R}{\lambda_{\rm TF}}\right) K_1\left(\frac{R}{\lambda_D}\right)}{ I_1\left(\frac{R}{\lambda_{\rm TF}}\right)
K_0\left(\frac{R}{\lambda_D}\right)}\right]^{-1}
\end{equation}
which is easily verified to be a function ranging from zero in the
limit $\lambda_D\ll\lambda_{\rm TF}$ to unity in the limit
$\lambda_D\gg\lambda_{\rm TF}$. We note that in the static limit,
water is highly polarizable and $\epsilon_3\sim 78 \epsilon_0$
while for silicon $\epsilon_1\sim 12 \epsilon_0$. {\highlight
Thus, for a fixed $n_0$ the ultimate sensitivity,
Eq.~(\ref{eq:bound}) with $\Gamma\sim 1$, requires a very short
Thomas--Fermi screening length, but at the same time one would
like to benefit from the $1/n_0$ dependence originating from
$G_0\propto n_0$.} We get $\lambda_D=\sqrt{\epsilon_3 k_B
T/2(Ze)^2 c_0}$ in the range of $\lambda_D\sim 1$ -- $100\,{\rm
nm}$ for at room temperature and ionic concentrations, $c_0$, for
typical physiological electrolytes. For
example~\cite{Heitzinger:2007} a realistic salt concentration
(Na$^+$Cl$^-$) of 150~mM yields a screening length of
$\lambda_D\sim 1\,{\rm nm}$ thus calling for nanowires with
densities supporting screening at the true nanometer scale. A
simple estimate of $\lambda_{\rm
TF}=\sqrt{\hbar^2\epsilon_1\pi^{4/3}/m^*e^2 n_0^{1/3}}$ also
yields $\lambda_{\rm TF}\sim 1\,{\rm nm}$ for a carrier
concentration of $10^{18}$\,cm$^{-3}$. {\highlight Taylor
expanding Eq.~(\ref{eq:Gamma_circle}) in the dilute carrier limit,
$\lambda_{\rm TF}\gg R$, we get
\begin{equation}\label{eq:Gamma_circle_dilute}
\Gamma\simeq  \frac{1}{2}\frac{\epsilon_1}{\epsilon_3}
\frac{\lambda_D}{R}\frac{
   K_0\left(\frac{R}{\lambda_D}\right)}{
   K_1\left(\frac{R}{\lambda_D}\right)}\left(\frac{R}{\lambda_{\rm
   TF}}\right)^2\propto n_0^{1/3}
\end{equation}
so that we arrive at the $n_0^{-2/3}$ scaling in
Eq.~(\ref{eq:bound}).}

\begin{figure}[t!]
\begin{center}
\epsfig{file=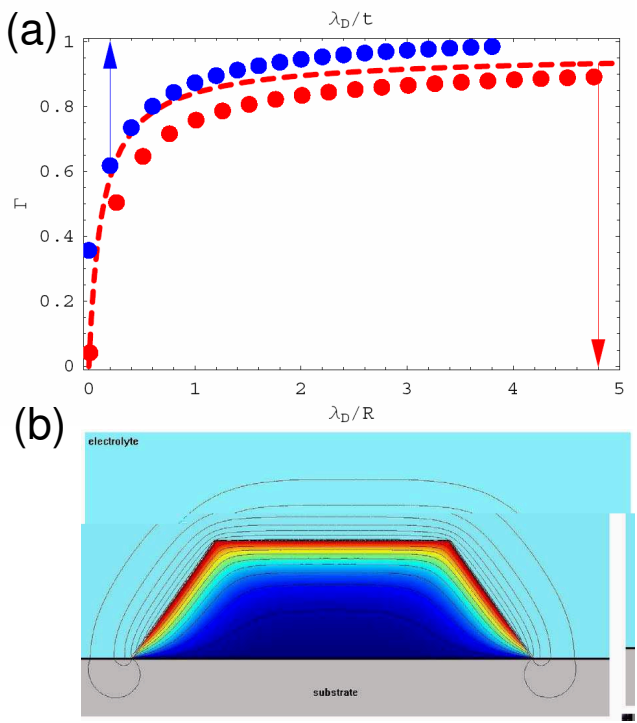, width=\columnwidth,clip}
\end{center}
\caption{(color online) (a) Sensitivity factor $\Gamma$ versus
Debye screening length $\lambda_D$. The lower trace: Cylindrical
nanowire with radius $R$ and $\lambda_{\rm TF}/R=0.02$. The data
points are the result of finite-element simulations with a finite
oxide layer thickness $\delta/R=0.005$. The dashed line shows
Eq.~(\ref{eq:Gamma_circle}). Upper trace: Finite element results
for the nanowire in (b) shown below. (b) Induced charge-carrier
density in a nanowire with trapezoidal cross section and
$\lambda_{\rm TF}/t=0.02$. Aspect ratio: $w/t=2$, etching defined
angle $\theta\sim 54,7^\circ$, oxide layer $\delta/t=0.02$.
Contours show the equipotential lines.} \label{fig2}
\end{figure}

Figure~\ref{fig2}a illustrates the dependence of the sensitivity
on the Debye screening in the electrolyte. The lower trace is for
a nanowire of circular cross section with $\lambda_{\rm
TF}/R=0.02$. The solid lines shows Eq.~(\ref{eq:Gamma_circle})
valid for a negligible oxide-layer thickness, i.e. $\delta\ll R$
while the data points are the result of finite-element simulations
(Comsol MultiPhysics) taking into account a finite oxide layer of
width $\delta/R=0.005$ and with $\epsilon_2=4$. As seen,
Eq.~(\ref{eq:Gamma_circle}) accounts well for the numerical exact
results. The upper trace shows finite element results for a
nanowire with a trapezoidal cross section with aspect ratio
$w/t=2$, an etching defined angle of $\theta\sim 54,7^\circ$
corresponding to the fabricated structure in
Ref.~\onlinecite{Stern:2007}, see Fig.~\ref{fig2}b. The oxide
layer has a thickness $\delta/t=0.02$ and for the Thomas--Fermi
screening we have $\lambda_{\rm TF}/t=0.02$ which is somewhat
stronger than for the circular case shown in the lower trace. Note
how the two curves have the same overall shape and dependence on
the Debye screening length, though the stronger Thomas--Fermi
screening for the upper case makes $\Gamma$ approach unity for
more moderate Debye screening lengths than in the lower case.
Fig.~\ref{fig2}b illustrates a typical distribution of the induced
charge-carrier density in the nanowire and the superimposed
contours show the equipotential lines. As expected the excess
carrier density is induced near the surface of the wire supporting
the surface-charge density $\sigma_S$.

Finally, let us discuss the prospects for sensing of point-like
charges located at a distance $\ell$ further away from the
conductor. Obviously, the additional Debye screening in the layer
of thickness $\ell$ (see Fig.~\ref{fig1}) will further reduce the
induced carrier density in the nanowire and in a simple picture
(neglecting curvature) we would qualitatively expect a reduction
proportional to $\exp(-\ell/\lambda_D)$. In the following we let
$N$ denote the average number of molecules absorbed on the wire of
length $L$ and we imagine that the chain-like functionalization
molecule supports a charged group, with charge $Q$, situated at a
distance $\ell$ from the surface. Smearing out these charges
results in an equivalent surface-charge density $\sigma_S =
NQ/[2\pi(R+\ell) L]$ at $r=R+\ell$. Solving the problem in
Eqs.~(\ref{eq:master}) and (\ref{eq:boundary}) for a finite $\ell$
we get $\Gamma\longrightarrow \Gamma_\ell\times \Gamma$ with
\begin{equation}
\Gamma_\ell \simeq 2\frac{R}{R+\ell}
\left[1+\sqrt{\frac{R}{R+\ell}}\:\exp\left(\frac{\ell}{\lambda_D}\right)\right]^{-1}.
\end{equation}
Here, we have used the large-argument exponential asymptotes for the Bessel functions. The
extra factor $0\leq \Gamma_\ell\leq 1$ illustrates the additional, close-to-exponential,
suppression by Debye screening when the charge is supported by a functionalization
molecule of length $\ell$. We note that in principle the $\lambda_D$ entering the
expression for $\Gamma_\ell$ could differ from the Debye screening length of the
electrolyte e.g. due to the surface functionalization.

In conclusion, we have used Thomas--Fermi and Debye--H{\"u}ckel
theory to formulate a simple screening model for surface-charge
sensing with conducting nanowires. The two screening mechanisms
are playing in concert and our model illustrates the non-trivial
interplay between the nanowire radius $R$, the Thomas--Fermi
Screening length $\lambda_{\rm TF}$, the Debye screening length
$\lambda_D$, and the length $\ell$ of the functionalization
molecules.

\emph{Acknowledgments.} We thank Jesper Nyg{\aa}rd,
Brian Skov S{\o}rensen, Per Hedeg{\aa}rd, Troels Markussen, and
Mogens H. Jakobsen for stimulating discussions.


\newpage

\end{document}